\documentclass[aps,
prl,
10pt,
floatfix,
superscriptaddress,
reprint,
twocolumn,
footinbib,
longbibliography]{revtex4-2}
\usepackage{xcolor}
\usepackage{blindtext}
\usepackage[T1]{fontenc}
\usepackage{graphicx}% Include figure files
\usepackage{soul}
\usepackage{dcolumn}% Align table columns on decimal point
\usepackage{svg}
\usepackage{bm}% bold math
\usepackage{mathtools}
\usepackage{amssymb}
\usepackage{amsmath}
\usepackage{amsthm}
\usepackage{listings}
\usepackage{bbold}
\usepackage{amsfonts}
\usepackage{appendix}
\usepackage{bm}
\usepackage[caption=false,subrefformat=parens]{subfig}
\usepackage{epsfig}
\usepackage{balance}
\usepackage[dvipsnames]{xcolor}
\usepackage{calc}
\usepackage[colorlinks,
linkcolor=blue,
anchorcolor=blue,
citecolor=blue,
urlcolor=blue]{hyperref}
\usepackage{orcidlink}
\usepackage{lipsum}
\usepackage{braket}
\usepackage{subfig}
\usepackage{verbatim}
\usepackage{cleveref}

\newcommand{\corr}[1] {{\color{black}{#1}}\color{black}}

\newcommand{\expSTMESR}{~\cite{Baumann2015Electron,
Natterer_Yang_nature_2017,Choi_Paul_natnano_2017,Willke_Paul_sciadv_2018,Yang2017Engineering,
Willke_Bae_science_2018,Bae2018Enhanced,Willke2019Tuning,
Willke2019Magnetic,Yang2019Tuning,yang_coherent_2019,Seifert_Kovarik_pr_2020,
Seifert_Kovarik_eabc_2020,Wang2023_universal,
Weerdenburg_Steinbrecher_2020,Steinbrecher_Weerdenburg_2020,Kim2022Anisotropic,Zhang2021Electron,kovarik_electron_2022,Kot2023Electric,Phark_double_resonance2023,Fe_driving_SH_2023,Wang2023Atomic,Zhang2023Influence,Bui2024All-electricalSpinb,kovarik_spin_2024,Wang2024Construction,czap2025magneticresonanceimagingsingle,Huang2025,greule2025spin}}
\begin{document}
%\title{Electric field control of the exchange field of a single spin impurity on a surface}

%% A small comment from me (Chris): what determines the transport is the DC bias (the difference of the chemical potentials) - the DC E-field is a *consequence* of changing the bias whilst keeping the tip height fixed, but it is not the *cause* of the exchange field in the more precise sense. I suggest using "DC bias" or "bias voltage" instead of DC E-field throughout.

%\title{Controlling the Exchange Field of Surface Spin Impurities using Electric Fields}
%\title{Controlling the Exchange Field of Surface Spin Impurities via DC bias}
\title{Controlling the Exchange Field of Surface Spin Impurities via DC Voltages}

\author{Xue Zhang}
\thanks{These authors contributed equally to this work.}
%\email{xuezhang@scut.edu.cn}
\affiliation{Spin-X Institute, State Key Laboratory of Luminescent Materials and Devices, Center for Electron Microscopy, South China University of Technology, Guangzhou 511442, China}
\affiliation{School of Microelectronics, South China University of Technology, Guangzhou 511442, China}
\author{Jose Reina-G\'alvez}
\thanks{These authors contributed equally to this work.}
\affiliation{Department of Physics, University of Konstanz, D-78457 Konstanz, Germany}
%\thanks{Previously at: Center for Quantum Nanoscience, Institute for Basic Science (IBS), Seoul 03760, Republic of Korea.}
\affiliation{Center for Quantum Nanoscience, Institute for Basic Science (IBS), Seoul 03760, Republic of Korea}
\affiliation{Ewha Womans University, Seoul 03760, Republic of Korea}
\author{Di{'}an Wu}
\affiliation{School of Microelectronics, South China University of Technology, Guangzhou 511442, China}
\author{Jan Martinek}
\affiliation{Institute of Molecular Physics, Polish Academy of Science, Smoluchowskiego 17, 60-179 Poznan, Poland}
\author{Andreas J. Heinrich}
\email{heinrich.andreas@qns.science}
\affiliation{Center for Quantum Nanoscience, Institute for Basic Science (IBS), Seoul 03760, Republic of Korea}
\affiliation{Department of Physics, Ewha Womans University, Seoul 03760, Republic of Korea}
\author{Taeyoung Choi}
\email{tchoi@ewha.ac.kr}
\affiliation{Department of Physics, Ewha Womans University, Seoul 03760, Republic of Korea}
\author{Christoph Wolf}
\email{wolf.christoph@qns.science}
\affiliation{Center for Quantum Nanoscience, Institute for Basic Science (IBS), Seoul 03760, Republic of Korea}
\affiliation{Ewha Womans University, Seoul 03760, Republic of Korea}

%+%%%%%%%%%%%%%%%%%%%%%%%%%%%%%%%%%%%%%%%%%%%%%%%%%%%%%%%%
\begin{abstract}

Recent advances in scanning \corr{tunneling microscopy have enabled quantum-coherent control of single surface spins via all-electric electron spin resonance (ESR). Such control requires magnetoelectric coupling, since spin resonance is a magnetic effect. We show that a magnetic tip induces a bias-dependent \textit{exchange field} on a localized Anderson impurity via virtual particle exchange with the magnetic lead. This field differs from Heisenberg exchange and can be tuned, reversed, or suppressed by the bias voltage. Our model reproduces bias-controlled resonance shifts for $S=1/2$ titanium atoms and Fe(II) phthalocyanine, enabling spin control via the exchange field and revealing the magnetoelectric mechanism behind all-electric ESR for spin-based quantum technologies.}

\end{abstract}
%+%%%%%%%%%%%%%%%%%%%%%%%%%%%%%%%%%%%%%%%%%%%%%%%%%%%%%%%%

\maketitle

{\emph{Introduction}--}
The manipulation and control of individual quantum spins on surfaces is critical for advancing quantum information processing and spintronic devices at the nanoscale~\cite{Chen2023Harnessing,Bi2023Recent}. While oscillating magnetic fields delivered via strip lines have traditionally been used to control the quantum states of spins, electric fields offer significant advantages, including faster switching speeds and more localized application through local gate electrodes~\cite{Nowack2007Coherent,Nozaki2012Electric, Hsu2016Electric,Pham2019Selective,Asaad2020Coherent,Liu2021Quantum,Gilbert2023On}. \corr{To achieve this, it is crucial to understand how local electric fields can be converted to effective magnetic fields which can act on the spin state~\cite{fert20245}.
%Despite the potential advantages, the exact magnetoelectric coupling mechanism remains elusive~\cite{fert20245}.
%This is particularly crucial for quantum computing applications, where rapid gate operations are essential to mitigate decoherence effects. Furthermore, electric field control could lead to more energy-efficient spintronic devices, benefiting low-power, high-density information storage and processing technologies~\cite{Nozaki2012Electric,Liu2019Electric, fert20245}. Despite these potential advantages, the impact of local electric fields on the quantum properties of surface spins remains elusive. 
In this context, the recently developed combination of electron spin resonance (ESR) and scanning tunneling microscopy (STM) provides an excellent tool to probe and manipulate single atomic and molecular spins with atomic scale precision using purely AC electric fields\expSTMESR. 
%~\cite{Baumann2015Electron,Yang2019Tuning,Zhang2021Electron,Wang2023Atomic, Wang2024Construction}. 
Despite the success of this experimental technique in studying the interaction between electric fields and the local magnetic moments of spins on surfaces, open questions about the exact mechanism leading to the magnetoelectric coupling mechanism in all-electric ESR remain~\cite{delgado2021review}. }

In this Letter, we focus on \corr{the bias-voltage dependent exchange interaction between a single electron spin localized in a magnetic adsorbate on a substrate and the magnetic tip of the STM. } When the magnetic tip is sufficiently close to the impurity, the interaction between the spin and the tip is dominated by an \textit{exchange interaction}, which leads to a shift in the resonance frequency \corr{when changing } the tip-atom distance~\cite{Yang2019Tuning, Willke2019Magnetic,Seifert_Kovarik_pr_2020, Seifert2021,kovarik_spin_2024}. \corr{To understand the origin of this exchange interaction, we focus on its peculiar \textit{bias dependence at fixed tip-atom distance}. 
In this situation, varying the DC bias yields a nonlinear resonance frequency shift, an inversion of the shift direction when the tip polarization changes sign, and a dependence of the shift on the angle between tip polarization and adsorbate spin.
We also observe a sign reversal of the frequency shift when crossing a vanishing-shift point at non-zero DC bias.
%We also observed a change of the sign shift and a vanishing shift at non-zero DC bias. 
%Originally, this bias voltage dependence resonance frequency shift was attributed to $g$-factor modulation~\cite{Seifert2021,Ferron2019,Willke2019Magnetic,Liu2019Electric}, since the exchange field was expected to remain constant at fixed tip-atom distance. 
Originally, this bias-dependent resonance frequency shift was mainly attributed to $g$-factor modulation~\cite{Seifert2021,Kot2023Electric,Ferron2019,Liu2019Electric}, as the exchange field was expected to remain constant at fixed tip–atom distance.%~\footnote{Reference~\cite{Kot2023Electric} additionally measured a DC bias dependence on the exchange field, but its physical origin remained unclear and was only suggested as possibly arising from $g$-factor modulation.}. However, this explanation is ruled out by magnetic moment measurements that hardly indicate any bias dependence, consistent with our density functional theory (DFT) calculations. 

We propose instead that the bias dependence originates entirely from the exchange field $B_\mathrm{exch}$, due to modulation of virtual particle exchange between the spin-polarized tip and the impurity spin.
%We propose that the bias-dependent lies fully in the exchange field $B_\mathrm{exch}$, originating from the modulation of the virtual particle exchange between the spin-polarized tip and impurity spin. 
Our model is distinctively different from established Heisenberg-like exchange and based on a single-orbital Anderson impurity coupled to magnetic leads, which, under time-dependent modulation of the exchange field by an AC bias voltage, has been suggested as the origin of the all-electrical ESR mechanism in STM~\cite{reinagalvez2025}. Here we will show that the component of the exchange-field parallel to the total magnetic field acting on the impurity spin is responsible for the frequency shift, which is related to spin-dependent energy renormalization observed in DC-transport in molecular Kondo quantum dots~\cite{Martinek2003, Pasupathy2004Kondo, Hauptmann2008Electric}. The good agreement between theory and experiment can alleviate the remaining doubts about the magnetoelectric coupling mechanism that drives ESR in the STM, pointing towards the exchange field as the most likely source.}

\begin{figure}
\includegraphics[width=\linewidth]{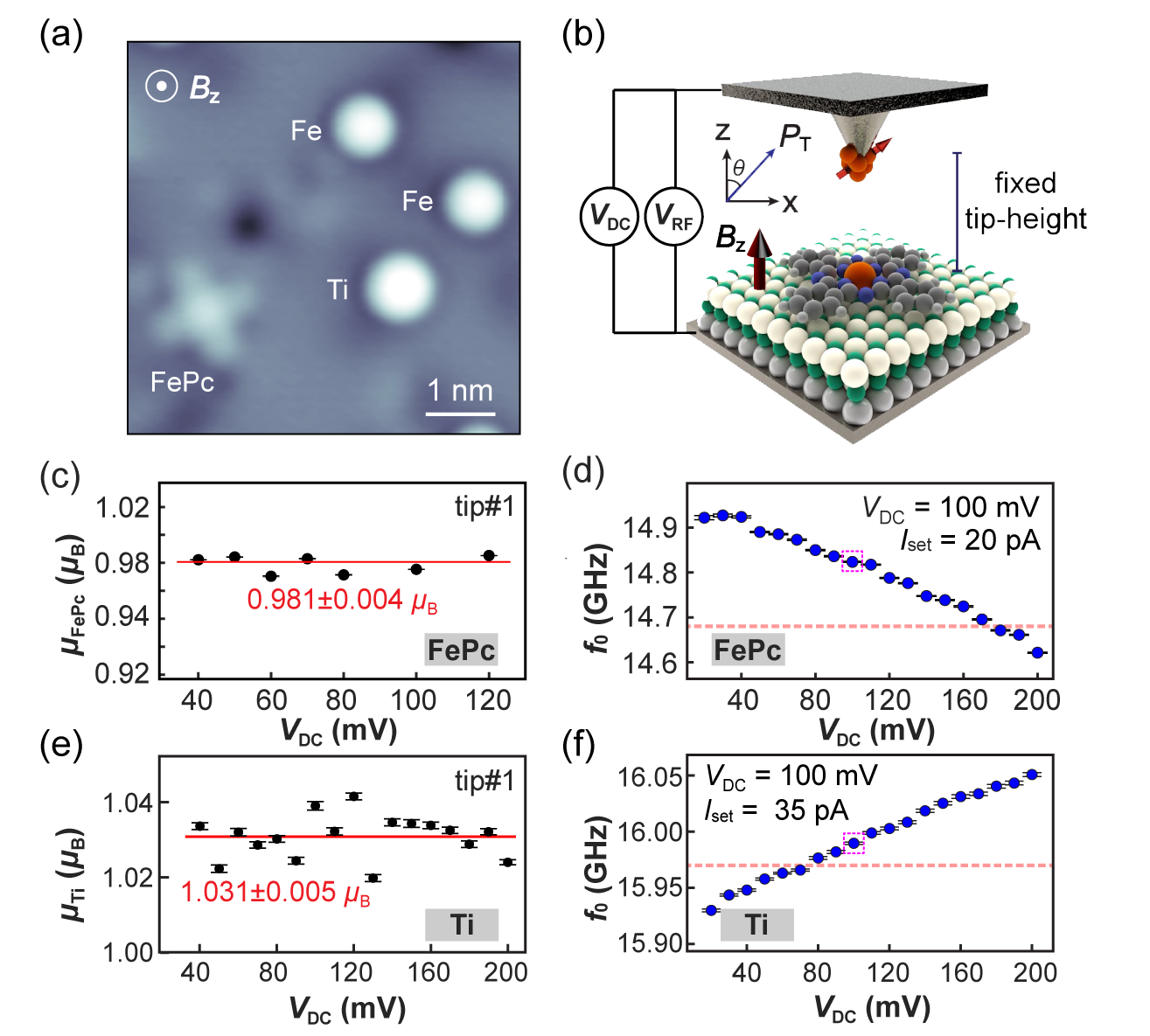}
\caption{{\bf ESR frequency ($f_0$) shift of an individual FePc molecule \corr{and a Ti atom } as a function of the DC bias.} (a) STM image of the Fe, Ti atoms and the FePc molecule co-adsorbed on the MgO/Ag(100) surface. Scanning parameters: $V_\text{DC} = 200$~mV, $I_\text{set} = 20$~pA. (b) Schematic diagram of the ESR set-up at the tunneling junction, integrated into a plate capacitor model. $\vec{P}_\mathrm{T}$ represents the spin-polarization of the tip, forming an angle $\theta$ with the out-of-plane direction. (c) Experimentally measured magnetic moments \corr{of FePc } at different $V_\text{DC}$. The red solid lines give the weighted average values of magnetic moments.  \corr{(d) } Corresponding resonance frequencies $f_0$ \corr{of FePc}.  \corr{The red dashed line indicates resonance frequencies calculated from DFT}. \corr{(e) and (f) show the same data for Ti. } The tip height was fixed with respect to the tunneling parameters of $V_\text{DC} = 100$~mV, $I_\text{set} = 20$~pA \corr{for FePc and $I_\text{set} = 35$~pA for Ti}. ESR conditions: $V_\text{RF} = 20$~mV \corr{for FePc and  $V_\text{RF} = 30$~mV for Ti}, $B_z = 560$~mT, $T = 1.8$~K.}
%(e) DFT-calculated $g_z$ for a FePc molecule as a function of an applied out-of-plane DC electric field. (f) Simulated $f_0$-shift as a function of $V_\text{DC}$ based on the $g_z$ modulation in (e).
\label{fig:1}
\end{figure}

{\emph{Experimental Results}--}
In our experiment we utilized two adsorbates, FePc molecules and Ti atoms (Fig.~\ref{fig:1}(a)), which have electron spin $S=1/2$ when adsorbed on a bilayer of MgO grown on top of an Ag(100) substrate~\cite{Zhang2021Electron,Yang2019Tuning}. The external magnetic field in this case was applied along out-of-plane direction ($B_{\mathrm{ext}}=B_z$) which, for a $S=1/2$, defines the spin quantization axis since the external magnetic field is much larger than the exchange field~\cite{reinagalvez2025,SM4}. Consequently, the resonance frequency is determined by \corr{the sum of external field and the exchange field component parallel to the quantization axis,}
\begin{equation}
   hf_0 = g\mu_\mathrm{B} (B_{\text{ext}} + B_{\text{exch}}^{\parallel}),
    \label{eq:Zeeman}
\end{equation}
\corr{with $h$ the Planck constant, $g$ the $g$-factor and $\mu_\mathrm{B}$ the Bohr magneton. The $g$-factor can be determined by the frequency response to the total magnetic field~\cite{SM1}. We note that the $g$-factor depends on the adsorbed atomic or molecular species and the local crystal field environment and can deviate significantly from the free electron value. } We performed ESR measurements on individual spins by adding a radio-frequency AC voltage ($V_\text{RF}$) in addition to $V_\text{DC}$ to the STM tip, as illustrated in Fig.~\ref{fig:1}(b). We follow the convention that $V_\text{DC}<0$ corresponds to probing the filled states of the sample.

One possible explanation of the aforementioned $V_{\mathrm{DC}}$ dependence of the resonance frequency is a bias dependence of $g$ in  Eq.~\eqref{eq:Zeeman}, i.e. $g=g(V_{\mathrm{DC}})$~\cite{Kot2023Electric,Liu2019Electric}. To estimate the $g$-factor variation in FePc, we measured the magnetic moment $\mu$ under different external fields, \corr{as deviations of $\mu$ from $1 \mu_\mathrm{B}$ imply a change of $g$ in Eq.~\eqref{eq:Zeeman}}. \corr{For FePc $\mu\approx 1 \mu_\mathrm{B}$~\cite{Zhang2021Electron, Chen2025FePc} and } remains largely constant across different $V_{\text{DC}}$ values (Fig.~\ref{fig:1}(c)) indicating that the $g$-factor does not significantly change within the bias range. Simultaneously, we observed a significant monotonic decrease in resonance frequency $f_0$ with increasing $|V_{\text{DC}}|$ \corr{in the same conditions } (Fig.~\ref{fig:1}(d)), ruling out \corr{a change of $g$-factor as dominant source of the frequency shift}~\footnote{We note that a precise determination of the $g$-factor from ESR-STM measurements is challenging due to competing influence of external and tip fields~\cite{Seifert2021}.} \corr{and calling for an alternative explanation. The lack of a strong electric-field dependence of the $g$-factor is further supported by our density functional theory } (DFT) calculations using a plate capacitor model that considers both electrostatic and geometric changes to the $g$-tensor~\cite{SMDFT}. \corr{DFT  predicts a negligible } frequency shift ($\sim$2 MHz, \corr{red dashed lines in Fig.~\ref{fig:1} (d)}) two orders of magnitude smaller than our experimental observations. This is consistent with previous studies of pentacene molecules on thick insulating layers using atomic-force microscopy, i.e. in the absence of a polarized tunneling current~\cite{Sellies2023Single}.

Further analysis also rules out DC bias-induced vertical displacements of the adsorbate as a significant contributor to the observed $f_0$ shift. By examining the tunneling $I$-$z$ curve of FePc at various $V_\text{DC}$ values, we calculated that maintaining a constant $f_0$ across our $V_\text{DC}$ range ($20$ to $200$~mV) would require a tip-adsorbate distance change of approximately 20~pm. This implies an assumed experimental displacement of $\sim$70~pm/(V/nm) (calculated with an estimated tip-sample distance of $\sim$0.6~nm~\cite{SM4}) Such a substantial vertical displacement should be easily detected by STM~\cite{soe2019low}; yet it was not observed in our experiments. Further supporting this conclusion, our DFT calculations predict that the mechanical displacement of FePc adsorbed on the MgO surface is only 0.44~pm/(V/nm) when subjected to a DC electric field. This suggests that the DC bias has negligible influence on the vertical adsorbate position in our $V_\text{DC}$ range and therefore does not contribute to the $f_0$ shift. 

To confirm the generality of the $V_\text{DC}$ induced resonance frequency shift, we studied individual Ti atoms for comparison. Similarly, the measured magnetic moments of Ti \corr{($\mu_{Ti}$) } show no evident dependence on the $V_\text{DC}$ with an average value of 1.031 $\mu_\text{B}$ (\corr{Fig.~\ref{fig:1}(e)})\corr{, in general agreement with previous determinations of the $g$-factor}~\cite{Kim2021Spin}. However, the resonance frequency increases by $\sim$150 MHz as $V_\text{DC}$ varies from 20 to 200 mV at fixed tip height (\corr{Fig.~\ref{fig:1}(f)}). This distinct behavior again rules out $g$-factor changes as the primary mechanism. DFT calculations \corr{performed in accordance with the calculations for FePc (red dashed line in Fig.~\ref{fig:1}(f)) show minimal $g_z$ variation with the DC electric field, } and predict a slight frequency decrease ($\sim$20 MHz within $\pm{600}$ mV) with more positive $V_\text{DC}$, contrary to experiments. Furthermore, the observed $\sim$150 MHz shift would require a tip adsorbate displacement of 50~pm for compensation~\cite{SM4} in the $V_\text{DC}$ range of $20$ to $200$~mV, which corresponds to $\sim$170 pm/(V/nm) and has not been experimentally observed. In contrast, DFT predicts only 0.096 pm/(V/nm) for the atomic displacement. These findings conclusively demonstrate that neither $g$-factor modulation nor adsorbate displacement can satisfactorily explain the observed frequency shifts.

%\chris{We further corroborate that these shifts are consistent across measurements obtained from different tips, since each tip can have slightly different apex geometries leading to different tip polarizations $\vec{P}_\mathrm{T}$.}

%As shown in Fig.~\ref{fig:2}(c) the resulting frequency shifts appear to be tip-dependent at first sight. However, it can be shown that the frequency shift is in fact near identical when rescaling the data for different tip polarizations (details of the rescaling can be found in \revision{SM} Sec. 3) as shown in Fig.~\ref{fig:2}(d).}

{\emph{Exchange field Model}--}
Given that changes of the $g$-factor and vertical displacement are insufficient to explain the significant resonance frequency shift, we propose that it is mediated by a DC bias-dependent exchange field.
The magnitude of $B_\text{exch}$ \corr{depends on } the probability of virtual electrons (or holes) exchange between the magnetic tip and the impurity and is not a mechanism exclusive to STM but generally appears in DC transport in magnetic tunneling junctions~\cite{Koenig2003,Braun2004Theory,Weymann2005Tunnel, Busz2023Hanle,Busz_martinek_2025}.

The resonance frequency shift $\delta f$ caused by the exchange field is 
\begin{equation}
   h\delta f = hf_0 - g\mu_\text{B} B_\text{ext} = g\mu_\text{B} B_\text{exch} \cos \theta = g\mu_\text{B} B_\text{exch}^\parallel.
    \label{eq:hdeltaf}
\end{equation}
Equation~\eqref{eq:hdeltaf} introduces the angle $\theta$ between the direction of the magnetization of the spin-polarized tip $\vec{n}_\mathrm{T}=\vec{P_\mathrm{T}}\left/\left|\vec{P_\mathrm{T}} \right|\right.$ and the external magnetic field $B_\text{ext}$. This angle $\theta$ indicates that only the projection of the exchange field onto the spin quantization axis $B_\text{exch}^\parallel$ contributes to the resonance frequency shift. Therefore, when $\theta=\pm 90^\circ$, the projection is zero and the resonance frequency shift vanishes~\footnote{In contrast, parallel alignment ($\theta=0^\circ, 180^\circ$) produces a maximum shift but no ESR signal as both the exchange field and the spin are parallel, causing, on resonance, no change in tunneling magnetoresistance~\cite{ReinaGalvez2021All,ReinaGalvez2023Many,reinagalvez2025}}. In the experiment $\theta$ appears in the homodyne detection and influences the asymmetric line shape of the ESR signal~\cite{Yang2017Engineering, Bae2018Enhanced, delgado2021}.

We now discuss the parameters that determine the exchange field. Under conditions typically found in low-temperature ESR-STM measurements, $\vec{B}_\text{exch}$ is given by~\cite{SM6,reinagalvez2025}:
\begin{equation}
  g\mu_\text{B} \vec{B}_\text{exch} = \frac{1}{2\pi} \left[ \gamma_\mathrm{T} P_\mathrm{T} \ln \left|\frac{eV_\text{DC}-\varepsilon-U}{eV_\text{DC}-\varepsilon}\right| \right] \vec{n}_T.
    \label{eq:exchB}
\end{equation}
 
\begin{figure}
\includegraphics[width=\linewidth]{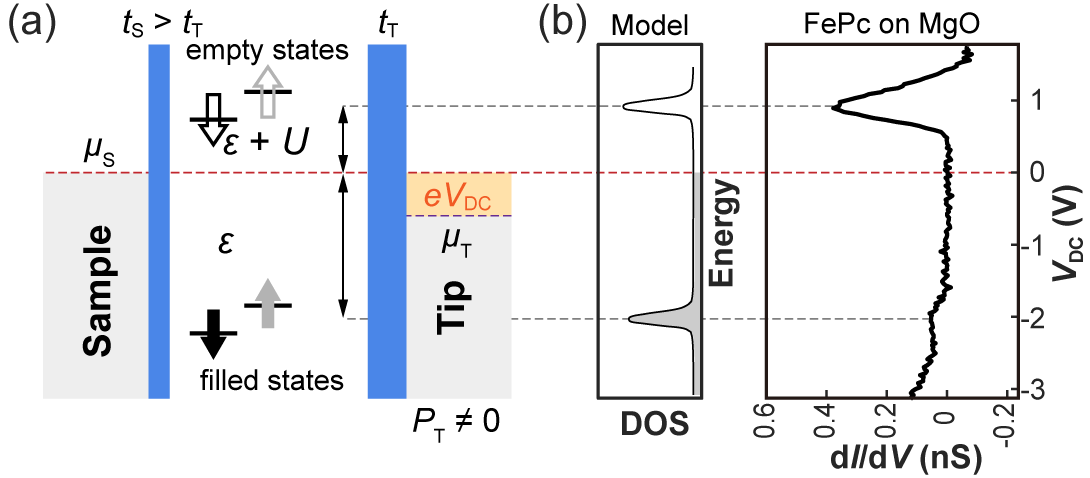}
\caption{{\bf Energy diagram and d$I$/d$V$ spectrum for SAIM.} 
(a) Energy-level diagram illustrating the key transport parameters governing the exchange field. 
The spin impurity splits into two states, $\downarrow$ (ground state) and $\uparrow$ (excited state), separated by the Zeeman energy. 
Filled and empty arrows denote singly occupied electron states and hole states, respectively. 
The applied DC bias is $eV_\text{DC}=\mu_S-\mu_T$. 
(b) Simulated density of states (DOS, left) and experimental d$I$/d$V$ spectrum (right) of a FePc molecule on MgO.}
\label{fig:2}
\end{figure}

\begin{figure}
\includegraphics[width=\linewidth]{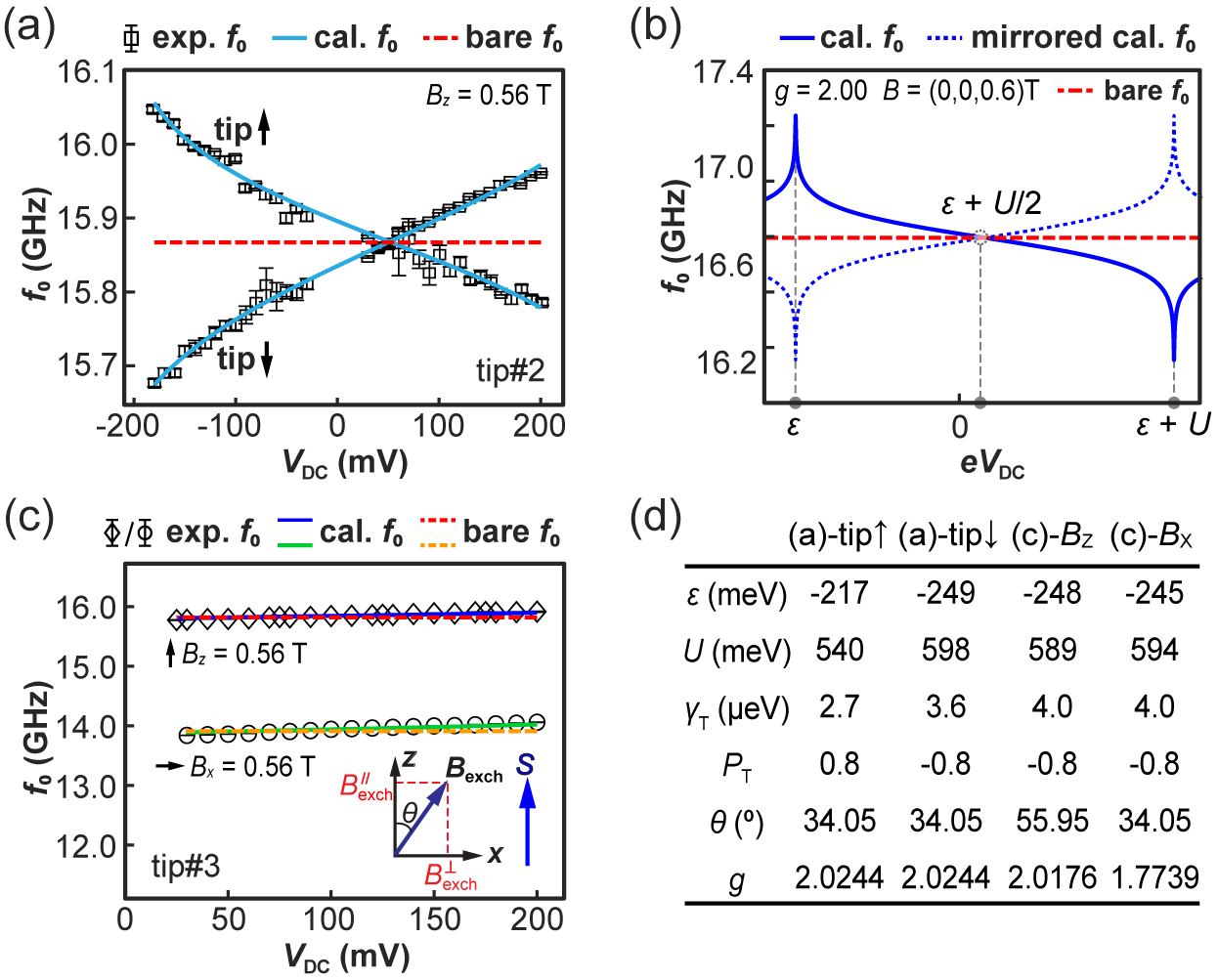}
\caption{{\bf Resonance frequency shift of a Ti atom measured by bistable tip.} (a) Experimentally measured (squares) and simulated (solid lines) frequency shift at varied $V_\text{DC}$ with a bistable tip showing two opposite signs of the spin-polarization (tip up, positive, and tip down, negative). 
In both (a) and (b), the red dashed line represent the bare resonance frequency set by external magnetic field and $g$-factor. The blue curves in (b) shows the resonance frequencies calculated using Eqs.~\eqref{eq:hdeltaf} and~\eqref{eq:exchB} with two opposing signs of the polarization (solid and dashed lines). 
(c) Experimentally measured (diamonds and circles), simulated (blue and green solid lines) and bare (red and orange dashed lines) frequencies vs $V_\text{DC}$ with the external magnetic field along the out-of-plane ($B_z$) and in-plane ($B_x$) direction, respectively. Inset: geometry of the exchange field indicating the homodyne angle $\theta$ and the parallel (perpendicular) projections $B_{\text{exch}}^{\parallel}$ ($B_{\text{exch}}^{\perp}$) onto the spin $\vec{S}$. (d) Parameters used for the simulations in (a) and (c). ESR experimental parameters: (a) $V_\text{RF} = 28$ mV, (b) $V_\text{RF} = 30$ mV. The tip height at different $V_\text{DC}$ was fixed by referring to the tunneling condition of (a) $V_\text{DC} = 100$ mV, $I_\text{set} = 20$ pA, (c) $V_\text{DC} = 100$ mV, $I_\text{set} = 40$ pA.}
\label{fig:3}
\end{figure}

Equation~\eqref{eq:exchB} encapsulates all relevant transport parameters, primarily depending on the ionization ($\varepsilon$) and charge ($\varepsilon+U$) energies of the SAIM schematically depicted in Fig.~\ref{fig:2}(a). To correlate the charging energies of the adsorbate with the SAIM levels, we show a representative d$I$/d$V$ spectrum of an individual FePc molecule on MgO, and compare it with the simulated density of states, Fig.~\ref{fig:2}(b)~\footnote{This interpretation relies on a change in the impurity’s charge state, $N \leftrightarrow N\pm1$~\cite{kumar2025theoryscanningtunnelingspectroscopy}.}.
%The exchange field vanishes when the Coulomb repulsion energy $U=0$, which demonstrates that the origin of the exchange field is a quantum many-body effect~\cite{Braun2004Theory,Weymann2005Tunnel,Martinek2005Gate,Busz2023Hanle}. 
\corr{The energies $\varepsilon$ and $\varepsilon+U$ determine the onset of  the non-linear region of the resonance frequency shift, since  $\vec{B}_\text{exch}$ increases logarithmically when the $eV_\mathrm{DC}$ approaches $\varepsilon$ or $\varepsilon + U$ (see Eq.~\eqref{eq:exchB} and Fig.~\ref{fig:3}(b))}. Crucially, Eq.~\eqref{eq:exchB} shows \corr{that $\vec{B}_\text{exch}$ is independent } from the resonance frequency, \corr{i.e from the Zeeman splitting, which is } negligible compared to \corr{$\varepsilon$ and $\varepsilon+U$ for the standard values of the magnetic field used here (cf. Refs.~\cite{reinagalvez2025,Braun2004Theory}), consistent with previous experimental findings~\cite{Seifert2021,Kot2023Electric} and in contradiction with a frequency shift mediated by the $g$-factor~\cite{SM9}}. \corr{Finally, we emphasize that the exchange field here is distinctively different from localized tip-spin models used to describe the interaction between tip and surface spin in previous works~\cite{Lado2017exchange, Yang2019Tuning} and provides a more natural description of the $V_{\mathrm{DC}}$-dependence. From the model point of view, it is the charge fluctuations of the Anderson model that provide this exchange field, which is absent in models that only describe spin fluctuations~\cite{SM5}.}

The DC bias in Eq.~\eqref{eq:exchB} is defined as the difference between the electrochemical potentials of the sample (S) and tip (T), $eV_\text{DC}\equiv \mu_\text{S}-\mu_\text{T}$, following the standard sign convention~\footnote{It is important to note that the sign convention in~\cite{reinagalvez2025} differs from the one used here. In~\cite{reinagalvez2025}, the DC bias is defined as $eV_\text{DC} \equiv \mu_\text{T} - \mu_\text{S}$ but this does not affect the overall conclusions of the model.}. Consequently, electrons move from tip to sample at positive applied bias. In the double-barrier tunneling of tip-impurity-substrate geometry in STM, asymmetric couplings lead to an asymmetric bias drop~\cite{Tu_X_double-barrier,ReinaGalvez2023Many}. Since the substrate coupling is stronger than that of the tip in typical STM setups, we assumed $\mu_\text{S} \approx 0$ (grounded), so $eV_\text{DC} = -\mu_\text{T}$~\footnote{Considering $\mu_\text{S} \neq 0$ in Eq.~\eqref{eq:exchB} will solely imply a redefinition of the ionization energy $\varepsilon$, which can be thought as a gating effect. However, this does not change the key conclusions from this model.}.

The exchange field depends exponentially on the tip-adsorbate distance through the coupling $\gamma_\text{T}$, while the spin-polarization of the ferromagnetic tip, $P_\text{T}$, affects the exchange field linearly. $P_\text{T}=0$ results in a vanishing exchange field, which underscores the importance of a polarized electrode for the $V_\text{DC}$-induced frequency shift~\footnote{We note that the exchange field does not depend on the coupling to the substrate, $\gamma_\text{S}$, as it is unpolarized.}. Since only the product of $\gamma_\text{T}P_\text{T}$ enters Eq.~\eqref{eq:exchB} we cannot determine them individually from a single measurement and we set $\left| P_\text{T} \right|=0.8$ following~\cite{J_Cuevas_C_Ast_ESR_theory_2024}. The apparent ferromagnetic or antiferromagnetic coupling between the tip and the magnetic impurity is determined by the sign of $P_\text{T}\cos\theta$. For example, when $V_\text{DC}>0$, $P_\text{T}<0$, and $\theta \in (0,180^\circ)$, Eq.~\eqref{eq:exchB} yields a positive exchange field, and the resonance frequency shifts toward larger values, a ferromagnetic behavior. Conversely, reversing the sign of $P_\text{T}$ leads to an antiferromagnetic coupling.

This can be experimentally demonstrated by analyzing bistable tips with two opposing spin-polarizations. Bistable tips infrequently fluctuate between two magnetic states on a timescale that is slow compared to the signal integration of the ESR-STM~\cite{Seifert2021, Singha2021, Yu2023Magnetic}. Consequently, these tips produce spectra exhibiting two resonances that shift in opposite directions as a function of $V_\text{DC}$, as illustrated in Fig.~\ref{fig:3}(a). More importantly, both directions intersect at non-zero $V_{\mathrm{DC}}$, representing a symmetry point where electron and hole-like processes exactly cancel the exchange field~\cite{SM6}. 

Our model accounts for this phenomenon by considering both positive and negative spin-polarizations $\pm \left| P_\text{T} \right|$, yielding fits that accurately reproduce the experiment. This measurement directly reveals the crossing point, providing a better estimate of $\varepsilon$ and $U$. Moreover, the calculated bare resonance frequency aligns well with this intersection, as shown by the red dashed line in Fig.~\ref{fig:3}(b). The crossing-point at finite bias also demonstrates the independence of the exchange field from the tunneling current, as a non-zero net exchange field can exist even at zero $V_\text{DC}$ and at zero current current. Crucially, the trend of opposing shifts could not be explained simply by a modulation of $g_z$ as the sign of the DC bias does not change when the polarization is inverted, enhancing the validity of our model.

{\emph{Effect of the tip's magnetization angle}--}
The change in the sign of the spin-polarization can also be interpreted as a rotation of the angle $\theta_z$ since maintaining a $P_\text{T}<0$ with orientation $\theta_z$ is equivalent to $P_\text{T}>0$ with an orientation of $180^\circ-\theta_z$, as Eqs.~\eqref{eq:hdeltaf} and~\eqref{eq:exchB} show. To illustrate the impact of $\theta$, it is helpful to compare the $f_0$ shift between in-plane and out-of-plane magnetic fields, as shown in Fig.~\ref{fig:3}(c). This data set was obtained on a Ti atom using the same spin-polarized tip and tip-atom distance. If the $P_\mathrm{T}$ forms an angle $\theta_x$ with the $B_x$ direction, it should also form an angle $\theta_z=90^\circ-\theta_x$ with $B_z$ (inset of Fig.~\ref{fig:3}(c)). This fact allows us to predict the relative orientation of $P_\mathrm{T}$ since from Eq.~\eqref{eq:hdeltaf} we can write:
\begin{equation}
      \tan \theta_x = \frac{hf_{0z}-g_z \mu_\text{B} B_\text{ext}}{hf_{0x}-g_x \mu_\text{B} B_\text{ext}}.
    \label{eq:tanthetax}
\end{equation}

This implies that $g\mu_\mathrm{B} B_\text{exch}^\parallel$ has no further dependence on the $g$-factor as the right hand side of Eq.~\eqref{eq:exchB} shows~\cite{Braun2004Theory,reinagalvez2025}. Our data revealed different slopes in the shift of the resonance frequency of a Ti atom at varying $V_\text{DC}$ with out-of-plane and in-plane magnetic fields (Fig.~\ref{fig:3}(c)). We achieved good agreement with an angle of $\theta_x = \corr{34}^\circ$~\cite{SM8}, resulting in nearly identical adjustable parameters ($\gamma_\text{T}$, $\varepsilon$ and $U$, see Fig.~\ref{fig:3}(d)) for both data sets. 
%\corr{For an analysis of the uncertainties in the parameters extracted from optimized simulations, see \revision{SM}, Secs.~5–6, Figs.~S9 and ~S10. The exact determination of the homodyne angle $\theta$ is given in \revision{SM}, Sec.~7, Fig.~S12.}
This consistency across different field orientations strongly supports the validity of our model, \corr{as piezoelectric displacement has been shown to lead to significantly different modulation of the $x$ and $z$-component of $g$~\cite{Ferron2019}.}

%and contradicts the $g$-factor modulation or a piezoelectric displacement as the origin of the frequency shift, since they do not possess any dependence on the magnetic field orientation as they both depend on out-of-plane electric field modulation alone. 
%However, the uncertainty of the fit is relatively large due to the lack of a strong nonlinear regime in the experimental data, as shown in the \revision{SM} (Sec. 5, Fig.~S8).

\corr{As displayed by the } list of the best-fit parameters used for the model in Fig.~\ref{fig:3}(d), the impurity parameters $\varepsilon$ and $U$ are consistent within the fitting uncertainties~\cite{SM7}.
%, with detailed analysis of fit quality and uncertainty range of the fitting parameters provided in \revision{SM} (Sec. 5, Fig.~S8). %Analysis of the four datasets shown in Fig.~\ref{fig:3} yields average values of $\varepsilon=-240\pm13$ meV and $U=581\pm23$ meV. This is further confirmed by a statistical analysis of different datasets measured on Ti (see~\revision{SM.} Figs.~S3 and S4), indicating that the exchange mechanism is highly reproducible with different tips.  
Accurate determination of these parameters requires measurements in the strongly non-linear regime of the resonance frequency shift, which may be challenging to access due to the low activation barriers for adsorbate hopping and diffusion.

\begin{figure}
\includegraphics[width=\linewidth]{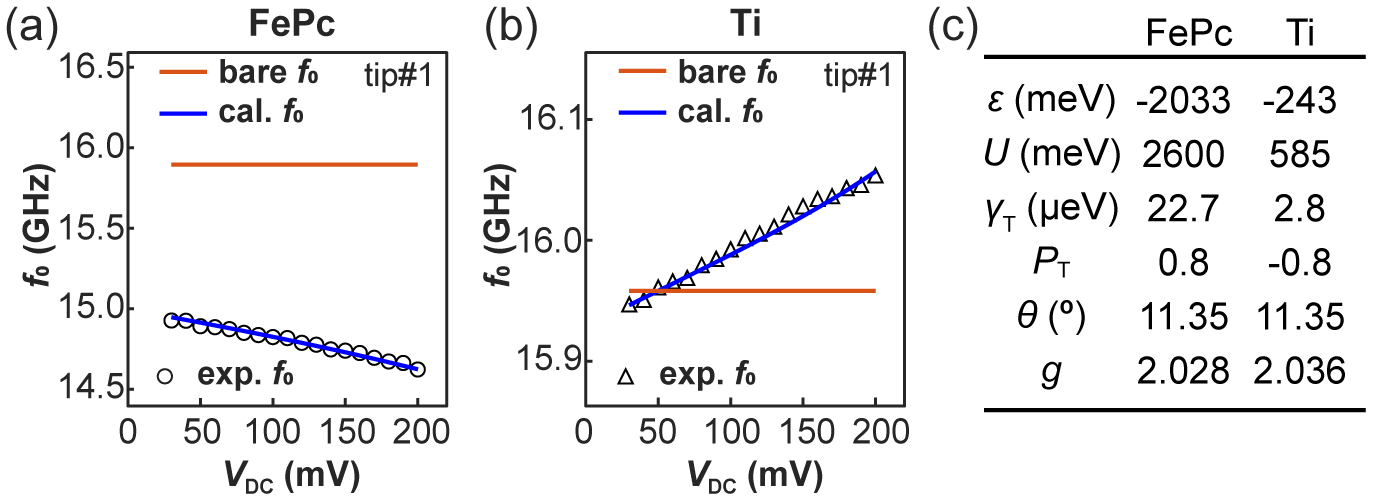}
\caption{{\bf Comparing the frequency shift of FePc and Ti measured with the same tip.} (a) and~(b) depict the experimental (circles and triangles) and simulated (solid lines) resonance frequencies as a function of $V_\text{DC}$  with the same tip for a FePc molecule and a Ti atom, respectively. The experimental values are extracted from Figs.~\ref{fig:1} and~\ref{fig:2}. (c) Parameters used in the transport simulations for FePc and Ti in (a) and~(b), respectively. \corr{Since the tip used here differs from that in Fig.~\ref{fig:3}, a different angle $\theta$ is chosen, approximately one third of $\theta_x$, which provides a reasonable fit of the ESR signal in terms of both linewidth and asymmetry.}}
\label{fig:4}
\end{figure}

{\emph{Comparison of different magnetic impurities}--} The exchange field model allows to differentiate between an FePc molecule and a Ti atom. Optimizing the simulation parameters to reproduce the ESR spectra of FePc and Ti measured with the same tip \corr{ (Fig.~\ref{fig:4}(a) and (b), respectively) shows consistent and reproducible values for each species, see Fig.~\ref{fig:4}(c)}. For Ti, we obtain values for $\varepsilon$, $U$, and $\gamma_\text{T}$ similar to those in Fig.~\ref{fig:3}(a) and (b). In contrast, FePc yields significantly different model parameters. When fixing $U = 2.6$ eV, based on the $dI/dV$ in Fig.~\ref{fig:2}(b), the model reproduces the frequency shift with $\varepsilon = -2.032$ eV which matches the negative charge resonance in Fig.~\ref{fig:2}(b)~\footnote{The energy alignment on the Ag/MgO surface places the unoccupied states closer to the Fermi level, making it easier to access via transport at positive bias. This could be a consequence of FePc on Ag/MgO being in an anionic state~\cite{Zhang2021Electron,Zhang2023Influence}.}. Allowing all parameters to vary in the fit gives slightly different $U$ and $\varepsilon$ from those in Fig.~\ref{fig:4}(c) for FePc~\cite{SM7}, but both fits indicate that FePc has a much higher Coulomb repulsion and ionization energy than Ti. \corr{Furthermore, while the lack of a strongly non-linear region in our frequency shift data makes a precise determination of $\varepsilon$ and $U$ difficult, our extracted values are in excellent agreement with more recent measurements on the same system which were performed with a $eV_\mathrm{DC}>\varepsilon+U$~\cite{greule2025spin}, thereby directly visualizing the non-linear region and confirming the validity of our model and the robustness of our parameter extraction. }
%The energy alignment on the Ag/MgO surface places the unoccupied states closer to the Fermi level, making it easier to access transport at positive bias. This could be a consequence of FePc on Ag/MgO being in an anionic state~\cite{Zhang2021Electron,Zhang2023Influence}.
%, and has direct implications for the measurement of the ESR signal, as it predicts lower Rabi rates for negative voltages compared to positive ones since $|\varepsilon|\gg\varepsilon+U$~\cite{reinagalvez2025,Urdaniz2025Transition}. 
\corr{Importantly, } the fitted $\gamma_\text{T}$ for FePc is substantially larger than that for Ti, suggesting a smaller tip-sample distance, consistent with experiments~\cite{Yang2017Engineering,Liu2019Electric,Willke2019Magnetic,Zhang2023Influence}.

\begin{figure}
\includegraphics[width=\linewidth]{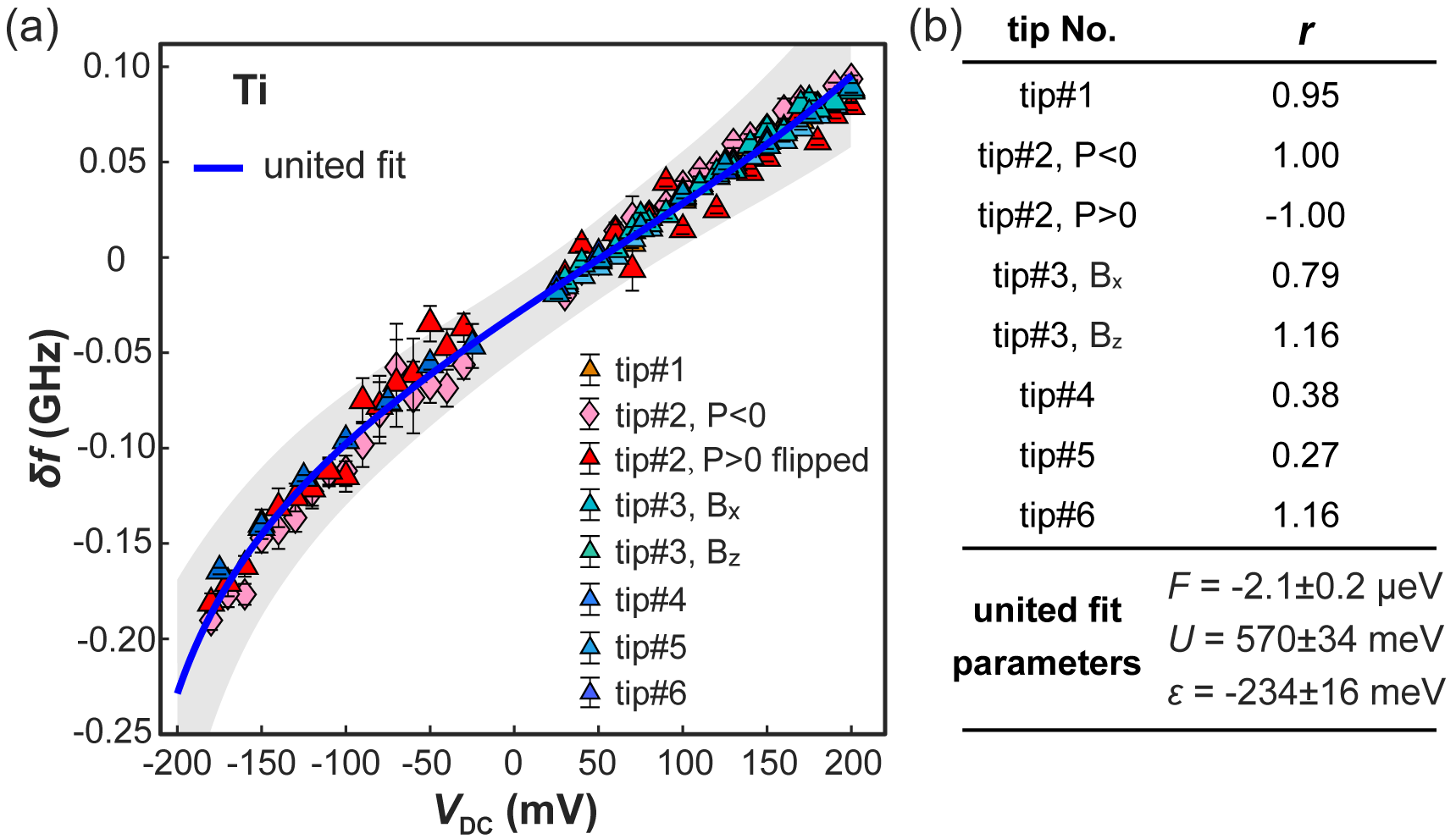}
\caption{\corr{{\bf Unified rescaled behavior of the resonance frequency shift of Ti described by the exchange-field model.} (a) Datasets from Fig.~\ref{fig:3}, Fig.~\ref{fig:4}, and additional ones (see Supplementary Material \cite{SM3}) are combined and rescaled to obtain a unified fit using a single set of adjustable parameters: $F = \gamma_\mathrm{T} P_\mathrm{T} \cos\theta$, $U$, and $\varepsilon$. The dataset acquired with a bistable tip of positive polarization is inverted. The grey shaded area indicates the overall uncertainty of the united fit. (b) Scaling factors $r$ used to align the datasets in (a) and the corresponding fit parameters with uncertainties indicated by the shaded area.}}
\label{fig:5}
\end{figure}

%\chris{Finally, we show that for each adsorbate the measured data follows a consistent frequency shift behavior. When combining all Ti data sets by calculating their frequency shift relative to their bare resonance ($\delta f$), which accounts for slight differences in the $V_\mathrm{DC}$-independent scaling factor  $F=\gamma_\mathrm{T} P_\mathrm{T}\cos(\theta)$ and external magnetic field in the individual measurements, they can be described by a single fit that reproduces the optimized parameters previously obtained in the individual fits; see Figs.~\ref{fig:5} and~S4(a), obtaining $U=570\pm 34$ meV, $\varepsilon=-234\pm 16$ meV and $F=-2.1\pm 0.2~\mu$eV. }
\corr{{\emph{Unified rescaled behavior}--} Finally, we show that the resonance frequency shift behavior is consistent across all measured data, Fig.~\ref{fig:5}(a). The relative frequency shift $\delta f$ of all Ti datasets falls on a single line after subtracting the external magnetic field and $g$-factor variations of each individual measurement, and rescaling the data to account for slight differences in the $V_\mathrm{DC}$-independent amplitude factor $F = \gamma_\mathrm{T} P_\mathrm{T} \cos\theta$. The resulting data is described by a single fit that reproduces the previously determined impurity parameters: $U = 570 \pm 34$ meV, $\varepsilon = -234 \pm 16$ meV, and an effective amplitude $F = -2.1 \pm 0.2~\mu$eV, as illustrated in Fig.~\ref{fig:5}(b). The scaling procedure and details of the analysis, together with the corresponding rescaled FePc data, are provided in~\cite{SM3}. These results demonstrate that all frequency-shift measurements share a common physical origin: the \textit{exchange field}.}

%\corr{The universality of our model is further supported by grouping the different Ti data sets discussed here into a single curve, as shown in Fig.~\ref{fig:5}. The resulting unified fit reproduces the individual results presented earlier and highlights that all frequency-shift measurements share the same physical origin: the \textit{exchange field}. The Ti data analyzed here appear to have been obtained under sufficiently similar conditions, which explains why the extracted fitted scaling factor is comparable across data sets. In contrast, additional data sets presented in \revision{SM, Sec.~5} requires rescaling to account for different measurement conditions, such as strong variations in polarization, coupling strength, or homodyne angle. By applying this procedure, all measurements are collapsed onto a single curve, see Fig.~S4(a). An analogous rescaling can be performed for the remaining FePc data, yielding a unified description as shown in Fig.~S4(b).}

{\emph{Conclusions}--}
In summary, we investigated \corr{how the DC bias influences } the electron spin resonance of individual molecules and metal atoms on a surface. Our \corr{transport } model reveals that the DC bias modulates the exchange field between a spin-polarized electrode and surface spins. This framework, combining exchange and external magnetic fields, captures the observed phenomena, including the difference between in-plane and out-of-plane magnetic fields, tip-angle effects\corr{, and the behavior of bistable magnetic tips through polarization inversion. Moreover, the frequency response to the DC bias enables the unambiguous identification of the two adsorbates observed in our experiments. This is supported by extensive statistical analysis across multiple STM tips, highlighting the potential of the exchange field as a tool for highly localized impurity spectroscopy. Altogether, our findings establish the DC bias as an effective parameter for tuning the exchange field in STM-based ESR experiments, overcoming potential limitations of the applicability of the ESR-STM technique. The presented model offers a practical framework for interpreting bias-dependent spin dynamics in nanoscale junctions and can be readily applied to related transport experiments. }

% The exchange field mechanism explains the observed resonance shifts and is consistent across different adsorbates and tip configurations.
%Finally, the exchange-field formalism developed can be applied to a variety of transport geometries with applications in spectroscopy, ,.. and quantum control.

%Finally, the exchange field model can not only describe all experimental observations in this paper but also fundamentally explain the mechanism of ESR in STM.

%Finally, the ability to tune local resonances using the exchange field from a polarized electrode opens new venues of quantum control in quantum-coherent nanoscience.

%Our findings demonstrate the potential of utilizing ESR-STM for characterizing ionization energies of surface spins and revealing fundamental physics of exchange fields stemming from spin-polarized electrodes in surface-spin systems, advancing atomic-scale quantum control through precise DC voltage manipulation.

{\emph{Acknowledgments}--}
All authors acknowledge support from the Institute for Basic Science under grant IBS-R027-D1. X.Z. acknowledges financial support from the National Natural Science Foundation of China (grant agreement No.~22202074, 22572060). J.M. received support from National Science Centre of Poland (Grant No.~2020/36/C/ST3/00539). J.R-G acknowledges support from the German Research Foundation [Deutsche Forschungsgemeinschaft (DFG)] under Projects No. 450396347. We also thank Nicol\'as Lorente, Hong Thi Bui, Soo-hyon Phark, Christian Ast, Piotr Kot, Arzhang Ardavan, Piotr Busz, Damian Tomaszewski, J\'ozef Barna\'s, and J\"urgen K\"onig for engaging discussions.

\section{Appendix}
{\emph{Sample preparation}--}
A commercial low-temperature STM system (Unisoku, USM1300) was used for all experiments. Sample preparation involved multiple steps: First, the Ag(100) substrate underwent cycles of Ar$^+$ sputtering and annealing to achieve atomically flat terraces. MgO film growth was accomplished by evaporating magnesium onto the clean Ag(100) surface at 400~$^\circ$C under an oxygen pressure of $1.1 \times 10^{-6}$~torr. 

FePc molecules, Ti and iron (Fe) atoms were deposited sequentially onto MgO/Ag(100). FePc molecules were deposited at room temperature, while Ti and Fe atoms were deposited onto the cold substrate. The sample was then cooled to a cryostat temperature of 1.8~K in the STM. The metal atoms and molecules are well-isolated from each other, with a low coverage of approximately 0.05 monolayers. The Ti atoms predominantly adsorb on the oxygen-oxygen bridge site in our work and are referred to simply as Ti. Fe atoms adsorb exclusively on the oxygen site and exhibit a shorter topographic height than Ti atoms. FePc molecules preferentially adsorb with their central Fe atom located on the oxygen site of the MgO lattice. \corr{All measurements were performed on Ti atoms and FePc molecules specifically chosen in defect-free regions of the surface. } We prepare spin-polarized tips by transferring several Fe atoms onto the tip apex with an applied external magnetic field of $\sim$0.6~T along the out-of-plane direction. All STM and ESR measurements were performed at 1.8~K with the external magnetic field applied out-of-plane or in-plane.

{\emph{ESR set-up}--}
The experimental setup for ESR measurements consisted of a signal generator (Keysight, E8257D) providing radio-frequency voltage $V_\text{RF}$. The DC bias ($V_\text{DC}$) was mixed with $V_\text{RF}$ through a bias tee and applied to the tip. The sample was grounded. For ESR signal detection, $V_\text{RF}$ was modulated at 95~Hz and measured using lock-in amplification. The tunneling current and ESR signal were detected by an electrometer connected to the sample. To maintain a constant tip-atom distance ($d$) as illustrated in the main text, we adjusted the setpoint current ($I_\text{set}$) correspondingly for each $V_\text{DC}$, thereby varying the DC bias (more details are provided in~\cite{SM2}). This method allows ESR measurements with active feedback using minimal gain, preventing unexpected thermal drift of the tip height. The ESR spectra are fitted with a Lorentzian to extract the resonance frequency $f_0$ for each applied $V_\text{DC}$. 

{\emph{Simulations}--}
DFT calculations were performed using plane-wave DFT with pseudopotentials implemented in the Quantum Espresso package~\cite{QE-2009, QE-2017}. Details of the input parameters are provided in~\cite{SMDFT}. 
\textit{Transport simulations} were carried out using our own method outlined elsewhere~\cite{ReinaGalvez2021All,ReinaGalvez2023Many,reinagalvez2025}. All solid curves in Figures~\ref{fig:3}-\ref{fig:5} are the result of transport simulations with optimized parameters for the impurity, tip angle and electrode-impurity couplings as listed in the respective panels. All simulations were performed at $T=1.8$ K and we use a fixed ratio of the tip and substrate couplings $\gamma_\mathrm{S}=3\gamma_\mathrm{T}$ where the unpolarized couplings are $\gamma_{\alpha}=2\pi|t_\alpha|^2 D_\alpha$, being $t_\alpha$ the tunneling amplitude and $D_\alpha$ the spinless density of states per electrode $\alpha$, with $\alpha=\text{T},\text{S}$ representing tip and sample, respectively. The density of states is assumed to be constant in the so-called wide-band limit. The spin-polarization of the ferromagnetic tip, defined from the spin-dependence density of states of the tip as $P_\text{T} \equiv (D_{T \uparrow}-D_{T \downarrow})/(D_{T \uparrow}+D_{T \downarrow})$, where $D_{T \sigma} = D_{T} \left(1/2 + \sigma P_{T} \right)$, was set to $P_\mathrm{T} = \pm 0.8$ following~\cite{J_Cuevas_C_Ast_ESR_theory_2024}. From the model perspective, a lower $P_\mathrm{T}$ can be compensated by a higher $\gamma_\mathrm{T}$, since both enter Eq.~\eqref{eq:exchB} as a product. However, the ESR signals are strongly influenced by the spin-polarization as well as the coupling to tip and substrate. More details of the transport simulations are provided in the Supplemental Materials~\cite{SM5,SM6,SM7,SM8}. %\corr{\textit{Rescaling of the data} accounts for the slightly different measurement conditions due to the stochastic nature of STM tip preparation. Extended data for Ti and FePc is presented in \revision{SM}, Sec.~3, Fig.~S4(a) and (b).}

%\corr{\textit{Rescaling of the data} accounts for the slightly different measurement conditions due to the stochastic nature of STM tip preparation, which influences $P_\mathrm{T}$, $\gamma_\mathrm{T}$, and $\theta$ as well as different external magnetic fields. However the exchange field can describe all frequency shifts $\Delta f$ after re-scaling the data. Extended data for Ti and FePc is presented in \revision{SM}, Sec.~3, Fig.~S4(a) and (b).}

% \corr{\textit{Rescaling of the data} In \revision{SM}, Sec.~3, Fig.~S4(a), three additional Ti data sets are completely accounted for in a unified plot after rescaling to compensate for substantial differences in $P_\mathrm{T}$, $\gamma_\mathrm{T}$, and $\theta$. A similar procedure was applied to FePc, adding three additional data sets to the one shown in the main text. All rescaled measurements for a given adsorbate fall onto a single curve (distinct for Ti and FePc), further demonstrating that the observed frequency shifts arise from the same physical mechanism, the \textit{exchange field}.}

\bibliographystyle{apsrev4-2} % Tell bibtex which bibliography style to use
\bibliography{ref}

\end{document}